\documentclass{iau} 
\usepackage{graphicx}

\title[Bow shocks in water fountain jets] 
{Bow shocks in water fountain jets}

\author[Gabor Orosz et al.]   
{Gabor Orosz$^{1,6}$, Jos\'e~F. G\'omez$^2$, Daniel Tafoya$^3$, Hiroshi Imai$^1$, Jos\'e~M. Torrelles$^4$, Ann Njeri Ngendo$^5$, and Ross A. Burns$^{6,1}$}

\affiliation{$^1$Kagoshima University, Kagoshima (Japan) \\ email: {gabor.orosz@gmail.com} \\[\affilskip]
$^2$IAA (CSIC), Granada (Spain) \\[\affilskip]
$^3$Chalmers, OSO, Onsala (Sweden) \\[\affilskip]
$^4$ICE (CSIC-IEEC), Barcelona (Spain) \\[\affilskip]
$^5$University of Nairobi, Nairobi (Kenya) \\[\affilskip]
$^6$JIVE, Dwingeloo (Netherlands)}

\pubyear{2017}
\volume{336}  
\jname{Astrophysical Masers: Unlocking the Mysteries of the Universe}
\editors{A. Tarchi, M.J. Reid \& P. Castangia, eds.}
\begin{document}

\maketitle

\begin{abstract}
We briefly introduce the VLBI maser astrometric analysis of IRAS~18043--2116 and IRAS~18113--2503, two remarkable and unusual water fountains with spectacular bipolar bow shocks in their high-speed collimated jet-driven outflows. The 22\,GHz H$_2$O maser structures and velocities clearly show that the jets are formed in very short-lived, episodic outbursts, which may indicate episodic accretion in an underlying binary system.
\end{abstract}

\firstsection 
\section{Introduction}

Post-AGB stars are important transitory objects as they are the link between AGB stars and PNe, two late stellar stages of evolution with strikingly different characteristics, yet barely separated in time. Explaining the formation and shaping of PNe, whose morphologies depart significantly from spherical symmetry, is one of the puzzling questions in late stellar evolution (see review in \cite{balickfrank2002}). It involves a sudden change in the mass-loss mode from spherical to bipolar/multipolar, which occurs on timescales of only a few hundreds of years. 

Among the most popular mechanisms that have been proposed to explain the presence of post-AGB axisymmetric lobes involve magnetic fields, torus-like equatorial density enhancements, and collimated bipolar jets. There is a growing consensus that these mechanisms cannot operate within single-star scenarios. Sensitive high-angular resolution observations have shown that many post-AGB objects exhibit high-speed collimated outflows, which seem to be carving bipolar cavities within slowly-expanding circumstellar envelopes (e.g., \cite{sahai2005}). It is thought that the bipolar structures seen in PNe are cavities created in the post-AGB phase and blown out by ionization and the fast wind from the hot central star.

In order to understand the origin of stellar high-speed collimated outflows, it is necessary to study in detail their kinematics, energetics, and physical conditions. Fortunately, there exists a group of post-AGB objects whose collimated outflows are traced by 22\,GHz H$_2$O maser emission which can be well characterized via interferometric and VLBI observations: the so-called Water Fountain stars (WF, see review in \cite{imai2007}).

\section{The water fountain IRAS 18043--2116}

The 22\,GHz H$_2$O masers of IRAS 18043--2116 were first mapped by \cite[Walsh et al. (2009)]{walsh2009} using the ATCA and covering a velocity range of $\sim$400 km~s$^{-1}$. They found a bipolar maser distribution with an outflow axis aligned close to the line of sight and confirming the source as a WF. \cite[P\'erez-S\'anchez et al. (2017)]{perezsanchez2017} detected radio continuum emission that they interpreted as the result of an ionizing shock front of a collimated high-velocity jet. In addition, \cite[Tafoya et al. (2014)]{tafoya2014} found that the source also hosts H$_2$O masers at 321\,GHz. The authors propose that these sub-mm H$_2$O masers arise in the same regions as their 22\,GHz counterparts (with a peak flux ratio of $\sim$1) and might be indicators of multiple jet launching events (see contribution by Tafoya et al., this volume).

\begin{figure}[tb]
\begin{center}
 \includegraphics[width=0.9\textwidth]{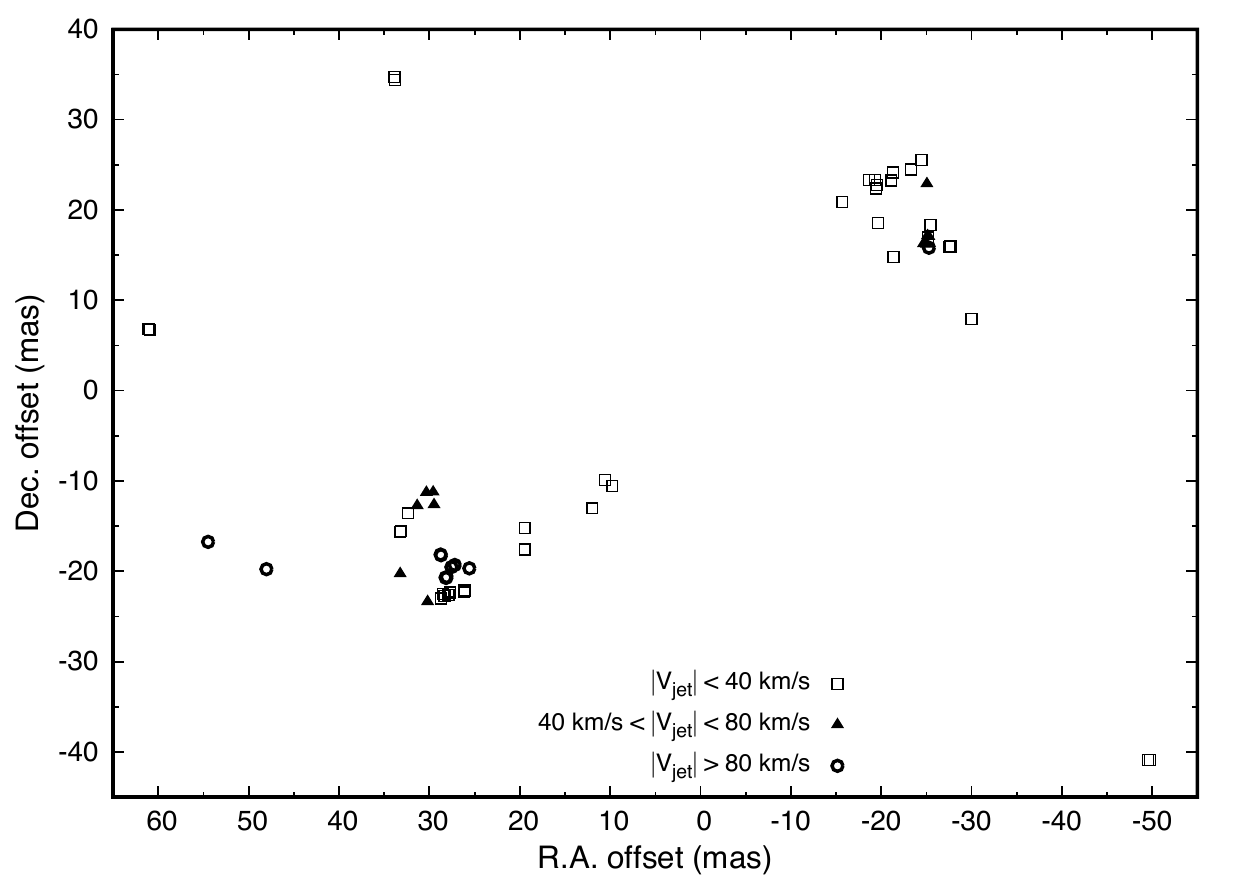} 
 \caption{Spatial distribution of 22~GHz H$_2$O masers towards the WF IRAS 18043--2116, derived from VLBA observations. The symbols are related to maser line-of-sight velocities relative to the systemic velocity of the star, i.e. $V_{jet} = V_{lsr, obs} - V_{star}$, with circles representing the fastest and squares the slowest masers in the outflows. The map origin is at the geometric center of the bipolar masers and the approximate location of the central object.}
   \label{iras18043map}
\end{center}
\end{figure}

To study the proposed jet scenarios, we mapped the region using archival 22\,GHz H$_2$O maser data taken with the VLBA (project BP150, \cite{day2011}). Figure~\ref{iras18043map} shows the detected maser features in the first two epochs. The masers trace bipolar arc-shaped structures along P.A.$\sim$128$^\circ$ that are blue- and redshifted relative to an LSR velocity of 87 km~s$^{-1}$ (\cite{deacon2004}). There are also three compact features that are spatially well removed from the bipolar structure, perhaps related to a relic AGB wind. In the eastern blueshifted lobe, the masers span line-of-sight velocities from $-$111 km~s$^{-1}$ to 78 km~s$^{-1}$, with the slower masers tracing an arc and the faster masers located mainly inside this structure. The redshifted lobe ranges from 94 km~s$^{-1}$ to 176 km~s$^{-1}$. Unfortunately at the time of the VLBA observations the true velocity extent of the source was not yet known, so the most redshifted masers above 245 km~s$^{-1}$ were not observed, and we cannot tell whether or not they also arise inside a slower arc.

We also measured proper motions, showing that the masers are expanding with an average outflow speed of $\sim$1 mas yr$^{-1}$. The kinematic age of the outflow is estimated to be $\lesssim$30 years. The maser proper motions can be well explained with a ballistic bow-shock model of a jet-driven outflow (see \cite{ostriker2001,lee2001}). H$_2$O masers can trace both the regions around the working surface of a collimated jet and the shocked shell at the interaction with the ambient material. It is therefore possible that the fastest masers (see circles on Fig.~\ref{iras18043map}) inside the eastern arc trace the tip of the high-speed jet, while the slower masers are related to a bow shock and a shell of ambient shocked gas. It is also possible that the faster masers are due to projection effects (as the jet axis is close to the line of sight), and the masers can be explained by the cavity model scenario introduced for W43A (\cite{chong2015}).

\section{The water fountain IRAS 18113--2503}

\begin{figure}[tb]
\begin{center}
 \includegraphics[width=0.9\textwidth]{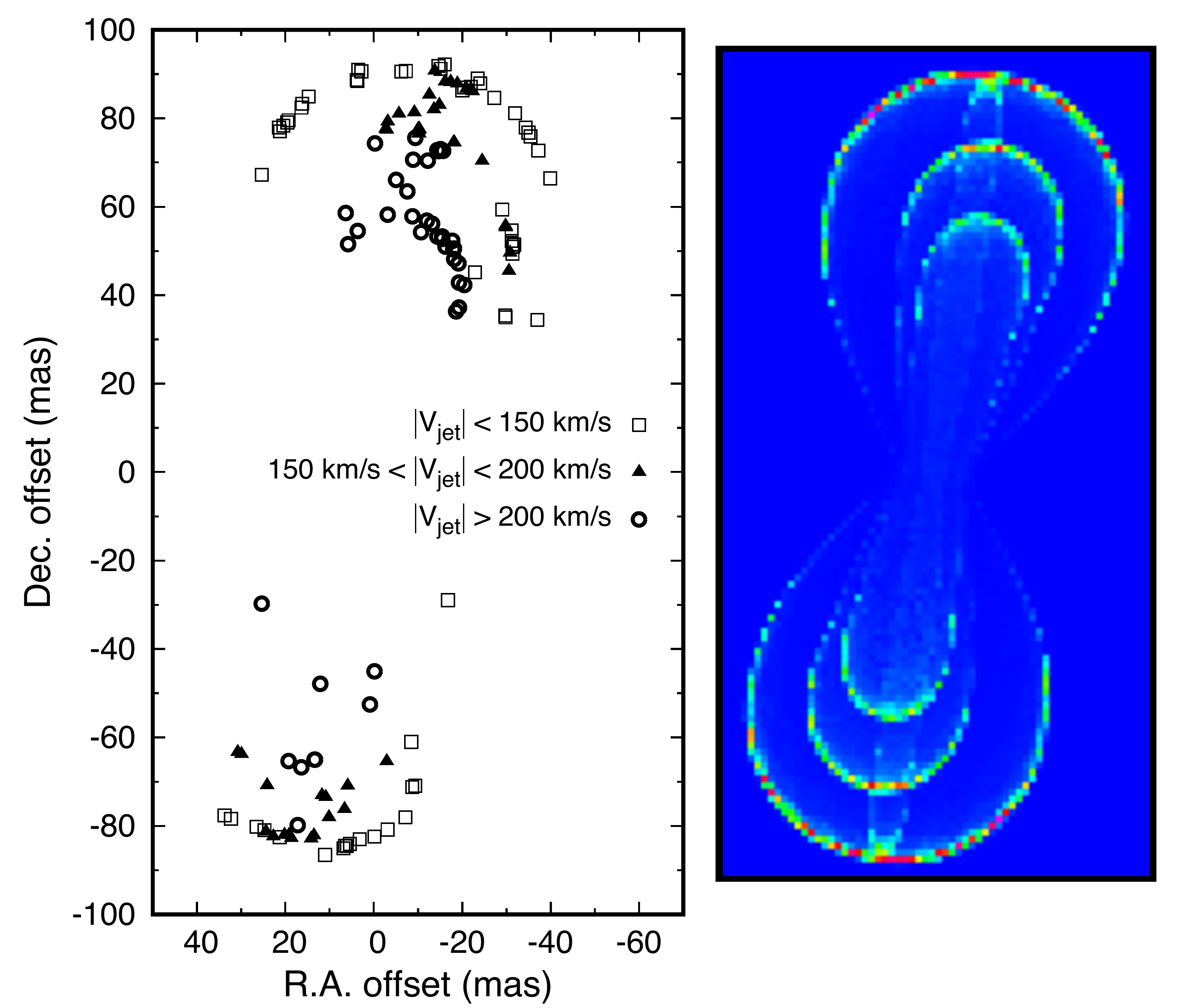}
 \caption{{\it Left:} Spatial distribution of 22~GHz H$_2$O masers towards the WF IRAS 18113--2503, derived from VLBA observations. The symbols are related to maser line-of-sight velocities relative to the systemic velocity of the star (same as Fig.~\ref{iras18043map}), with circles representing the fastest and squares the slowest masers in the outflows. The map origin is at the geometric center of the masers and the approximate location of the central object. {\it Right:} A simple rendered 3D morpho-kinematic ``Shape'' model made up of three bipolar lobes and a central conical jet. The pixel values have an arbitrary scale and are proportional to the calculated column density.}
\label{iras18113map}
\end{center}
\end{figure}

IRAS 18113--2503 was confirmed by \cite[G\'omez et al. (2011)]{gomez2011} to be a post-AGB WF with 22\,GHz H$_2$O masers spanning a very large 500 km~s$^{-1}$ line-of-sight velocity range. The masers were found to be in two spatially separated clusters, with each having a high velocity dispersion of approximately 170 km~s$^{-1}$. In order to characterize the physical size and 3D velocity field of IRAS 18113--2503, we started VLBI astrometric campaigns with the VLBA (project BG231) and VERA (project VERA13-85) arrays, and measured the proper motions and annual parallax of the H$_2$O masers. The left side of Fig.~\ref{iras18113map} shows the spatial distribution of the detected masers in the VLBA experiments. We found that the source is located at $\sim$12 kpc towards the Galactic Centre in the Galactic thick disk, and that the masers trace a nested axisymmetric bow shock structure along a position angle of $\sim$168$^\circ$, with a clear velocity gradient across the different arcs. Analyzing the 3D maser motions, we find that the jets slow down and get less collimated further from the star, from $\sim$270 km~s$^{-1}$ for the innermost jet to $\sim$140 km~s$^{-1}$ in the outermost part. 

A possibility is that IRAS 18113--2503 hosts an episodic jet and the H$_2$O masers trace very short-lived, episodic outbursts. The right side of Fig.~\ref{iras18113map} illustrates the rendering of a simple 3D morpho-kinematic model using the ``Shape'' program (\cite{steffen2011}), with parameters derived from the maser maps. Here we assume a high-speed conical jet and three axisymmetric lobes with slightly different orientations, caused by a possible precessing motion in the jet. Shape uses LTE radiative transfer codes for rendering, so the amplitudes cannot be directly compared to the masers. However, the model highlights the areas with the highest column density and as such the preferential places of the shock-excited H$_2$O masers. The goal is to visualize the 3D patterns that might be responsible for the complicated maser distribution on the sky.

In an episodic jet scenario, the more recent ejections could be faster and more collimated due to previous ejections sweeping up material and thus allowing the new ejections to expand more freely. It is also possible that the slowing shock fronts are not tracing three ejections but only the progression of a single ejection through an inhomogeneous medium, that could be associated with a layered relic AGB circumstellar envelope. However, it is interesting to note that while the three bow shocks are not spaced equally, the kinematic age difference between them are about 10 years and equal to within a year. Such episodic behavior might indicate that the underlying driving source of the bipolar mass ejections is a binary system with a period on the same timescale.

\section*{Acknowledgements}
GO acknowledges the support of the Monbukagakusho:MEXT scholarship, Kagoshima University, the Joint Institute for VLBI ERIC and the Konkoly Observatory.

\end{document}